\documentclass[a4paper,fleqn,usenatbib]{mnras}
\usepackage[T1]{fontenc}
\usepackage{ae,aecompl}
\usepackage{graphicx}	
\usepackage{amssymb}
\usepackage{amsmath}	

\newcommand{\HI}{HI}

\title[HI-mass halo-mass relation]{Constraints on the evolution of the
  relationship between \HI{} mass and halo mass in the last 12 Gyr}

\author[Padmanabhan and Kulkarni]{{Hamsa Padmanabhan$^{1}$\thanks{Email: hamsa.padmanabhan@phys.ethz.ch}
    and Girish Kulkarni$^{2}$\thanks{Email: kulkarni@ast.cam.ac.uk}} \\
  $^{1}$ Institute for Astronomy, ETH Zurich,
  Wolfgang-Pauli-Strasse 27, CH-8093 Z\"{u}rich, Switzerland\\
  $^{2}$ Institute of Astronomy and Kavli Institute for Cosmology,
  University of Cambridge, Madingley Road, Cambridge CB3 0HA, UK}

\date{Accepted ---. Received ---; in original form ---}

\pubyear{2015}

\begin{document}
\label{firstpage}
\pagerange{\pageref{firstpage}--\pageref{lastpage}}
\maketitle

\begin{abstract}
  The neutral hydrogen (\HI{}) content of dark matter haloes forms an
  intermediate state in the baryon cycle that connects the hot
  shock-heated gas and cold star-forming gas in haloes.  Measurement
  of the relationship between \HI{} mass and halo mass therefore puts
  important constraints on galaxy formation models.  We combine radio
  observations of \HI{} in emission at low redshift ($z\sim 0$) with
  optical/UV observations of \HI{} in absorption at high redshift
  ($1<z<4$) to derive constraints on the evolution of the HI-mass
  halo-mass (HIHM) relation from redshift $z=4$ to $z=0$.  We find
  that one can model the HIHM relation similar to the stellar-mass
  halo-mass (SHM) relation at $z \sim 0$. At $z=0$, haloes with mass
  $10^{11.7}$ M$_\odot$ have the highest \HI{} mass fraction ($\sim
  1\%$), which is about four times smaller than their stellar mass
  fraction.  We model the evolution of the HIHM relation in a manner
  similar to that of the SHM relation.  Combining this
  parameterisation with a redshift- and mass-dependent modified
  Navarro-Frenk-White (NFW) profile for the \HI{} density within a
  halo, we draw constraints on the evolution of the HIHM relation from
  the observed \HI{} column density, incidence rate, and clustering
  bias at high redshift.   We compare these findings with
    results from hydrodynamical simulations and other approaches in
    the literature, and find the models to be consistent
    with each other at the 68\% confidence level.
\end{abstract}


\section{Introduction}

Understanding the evolution of neutral hydrogen (\HI{}) in dark matter
haloes is important for models of galaxy formation
\citep{somerville2015, blanton2009, barkana2016}.  The \HI{} content
of dark matter haloes forms an intermediate state in the baryon cycle
that connects the hot shock-heated gas and star-forming molecular gas
in haloes \citep{2010ApJ...718.1001B, 2010MNRAS.409..515F,
  2012ApJ...753...16K}.  Constraints on \HI{} in galaxies therefore
reveal the role of gas dynamics, cooling, and regulatory processes
such as stellar feedback and gas inflow and outflow in galaxy
formation \citep{prochaska09, 2011MNRAS.414.2458V,
  2015MNRAS.447.1834B, 2015MNRAS.451..878K, 2016MNRAS.456.1115B}.
\HI{} also traces environmental processes like satellite quenching,
tidal interactions and ram-pressure stripping
\citep{2012MNRAS.427.2841F, 2012MNRAS.424.1471L, 2013MNRAS.429.2191Z,
  lagos2014}.  The average HI mass content of dark matter haloes can
be expressed as an HI-mass halo-mass (HIHM) relation.

At low redshifts ($z \sim 0$), constraints on \HI{} in galaxies are
derived from the observations of the 21~cm emission line of hydrogen
in large-area blind galaxy surveys like the \HI{} Parkes All Sky
Survey \citep[HIPASS,][]{meyer2004} and the Arecibo Fast Legacy ALFA
survey \citep[ALFALFA,][]{giovanelli2005}, which provide measurements
of the mass function and clustering of \HI{}-selected galaxies. There
are also targeted surveys such as The \HI{} Nearby Galaxy Survey
\citep[THINGS,][]{walter2008}, the Galex Arecibo SDSS Survey
\citep[GASS,][]{catinella2010}, and the Westerbork \HI{} survey of
Spiral and Irregular Galaxies \citep[WHISP,][]{vanderhulst2001}, which
focus on a smaller number of resolved galaxies. Efforts are also
currently underway to constrain the density and clustering of \HI{}
using intensity mapping without resolving individual galaxies
\citep{chang10, masui13, switzer13}. In the future, current and
upcoming facilities such as MeerKAT \citep{jonas2009}, the Square
Kilometre Array \citep[SKA,][]{2015aska.confE..19S} and its
pathfinders, and the Canadian Hydrogen Intensity Mapping Experiment
\citep[CHIME,][]{2014SPIE.9145E..22B}, will provide unprecedented
insight into the evolution of the cosmic neutral hydrogen content
across redshifts.

Unfortunately, the intrinsic faintness of the 21~cm line and the
limits of current radio facilities hamper direct detection of \HI{}
from individual galaxies at redshifts above $z \sim 0.1$. Spectral
stacking has been used to probe the \HI{} content of undetected
sources out to redshifts $z \sim 0.24$ \citep{lah07, lah2009, rhee13,
  delhaize13}. At higher redshifts, therefore, constraints on the
distribution and evolution of HI in galaxies come chiefly from high
column density Lyman-$\alpha$ absorption systems (Damped
Lyman-$\alpha$ Absorbers; DLAs) with column density
$N_\mathrm{HI}>10^{20.2}$~cm$^{-2}$ in the spectra of bright
background sources such as quasars.  DLAs are the main reservoir of HI
between redshifts $z\sim 2$--$5$, containing $> 80 \%$ of the cosmic
HI content \citep{wolfe1986, lanzetta1991, gardner1997, prochaska09,
  rao06, noterdaeme12, zafar2013}.  At low redshift, DLAs have been
found to be associated with galaxies \citep{lanzetta1991} and to
contain the vast majority ($\sim 81\%$) of the \HI{} gas in the local
universe \citep{zwaan2005a}.  At high redshift, the kinematics of DLAs may
support the hypothesis that they probe HI in large rotating disks
\citep{1997ApJ...487...73P, 2001MNRAS.326.1475M, 2015MNRAS.447.1834B} or proto-galactic clumps \citep{haehnelt1998}.
The three-dimensional clustering of DLAs \citep{fontribera2012} points
to DLAs being preferentially hosted by dark matter haloes with mass $M
\sim 10^{11} M_{\odot}$ at redshift $z \sim 3$.

Semi-analytical models and hydrodynamical simulations have provided
clues towards the evolution of \HI{} in galaxies and its relation to
star-formation, feedback and galaxy evolution \citep{dave2013,
  duffy2012, lagos2011, obreschkow2009a, nagamine2007, pontzen2008,
  tescari2009, hong2010, cen2012, fu2012, kim2013, bird2014,
  popping2009, popping2014, eagle2016, kim2016,
  martindale2016}. {Semi-analytical methods \citep[e.g.,][]{berry2014, popping2014, somerville15} typically reproduce the
\HI{} mass functions and the \HI{}-to-stellar-mass scaling relations
found in low-redshift \HI{} observations and DLA observables.}
Simulation techniques have also been used to model DLA populations at
higher redshifts \citep{pontzen2008} and their relation to galaxy
formation and feedback processes \citep{bird2014, rahmati2013,
  rahmati2014}.  Hydrodynamical simulations suggest that DLAs are
hosted in haloes with mass $10^{10}$--$10^{11} h^{-1}$ M$_\odot$
\citep[e.g.,][]{bird2014}.  In the presence of strong stellar
feedback, these simulations can reproduce the observed abundance and
clustering of DLAs but end up having an excess of \HI{} at low
redshifts ($z<3$).

Analytical techniques offer complementary insight into the processes
governing the \HI{} content of dark matter halos. Analytical methods
have been used for modelling 21~cm intensity mapping observables,
particularly the \HI{} bias and power spectrum
\citep{marin2010,wyithe2010, sarkar2016} as well as DLAs \citep{haehnelt1996,
  haehnelt1998,barnes2009, barnes2010, 2013ApJ...772...93K,
  barnes2014}. These models use prescriptions for assigning \HI{} mass
to dark matter halos as inputs to the model, either directly or in
conjunction with cosmological simulations \citep{bagla2010, marin2010,
  gong2011, guhasarkar2012}. In \citet{hptrcar2016}, the 21-cm- and
DLA-based analytical approaches are combined towards a consistent
model of \HI{} evolution across redshifts. It is found that a model
that is consistent with low-redshift radio as well as high-redshift
optical/UV observations requires a fairly rapid transition of \HI{}
from low-mass to higher-mass haloes at high redshifts. A more complete
statistical {data-driven} approach \citep{2016arXiv160701021P}
constrains the HIHM relation using low- and high-redshift
observations {in a halo model framework}.

An essential ingredient in analytical techniques is therefore the HIHM
relation.  In this paper, we employ the technique of abundance
matching to quantify the observational constraints on the HIHM
relation in the post-reionization Universe. Abundance matching has
been widely used to describe the relation between the stellar mass of
galaxies and the mass of their host dark matter halos \citep{vale2004,
  vale2006, conroy2006, behroozi2010, guo2010, shankar2006,
  moster2010, moster2013}.  The basic assumption involved is that
there is a monotonic relationship between a galaxy property (say,
stellar mass or galaxy luminosity) and the host dark matter halo
property (say, the host halo mass). In its simplest form, abundance
matching involves matching the cumulative abundance of galaxies to
that of their (sub)haloes, thereby assigning the most luminous
galaxies to the most massive haloes. The mapping between the
underlying galaxy property and the host halo mass can be derived from
this. {A key feature of this approach is that being completely empirical\footnote{{A caveat is that the halo mass function being used is theoretical, and the assumption of matching the most massive haloes is involved.}}, it is free from the uncertainties involved in physical models of \HI{} and galaxy evolution. {It is therefore a complementary analysis to forward modelling techniques, including semi-analytical models  and hydrodynamical simulations.}

The \HI{} mass function \citep{rao1993} is the radio equivalent of the
optical luminosity function in galaxies and is an important
statistical quantity in the observations of gas-rich galaxies. It
measures the volume density of \HI{}-selected galaxies as a function
of the \HI{} mass and simulations suggest that its shape is a more
sensitive probe of some aspects of galaxy formation physics than the
galaxy luminosity function \citep{kim2013}. At low redshifts, the
\HI{} mass function is fairly well-constrained over four decades in HI
mass \citep{zwaan05, martin10}. \citet{papastergis2013} constrained
the HIHM relation at low redshift using ALFALFA data and found that
the observed clustering of HI was reproduced well by this approach.  In
this work, we describe the results of abundance matching \HI{} mass to
dark matter halo mass using the low-redshift radio observations of the
\HI{} mass function \citep{zwaan05, martin10} and then evolve the
relation using the complementary information available through DLA
measurements at high redshift. The combination of the radio data at
low redshifts and DLA observations at higher redshifts constrains a
multi-epoch \HI{}-halo mass relation with the available data. We also compare how the results from this approach are consistent with those from studies in previous literature.

The paper is organized as follows. In Section~\ref{sec:abmatch}, we
detail the abundance matching technique and apply it to three
low-redshift HI mass function measurements.  We also combine the
resultant HIHM relation with the stellar-mass halo-mass (SHM) relation
to discuss the HI-to-stellar-mass ratio in low-redshift galaxies. In
Section~\ref{sec:observations}, we extend the low-redshift HIHM
relation to higher redshifts using measurements of DLA column density
distribution and clustering.  We compare the relation so derived with
other \HI{} models in the literature, and conclude in
Section~\ref{sec:conc}.

\section{HIHM relation at low redshift}
\label{sec:abmatch}

\begin{figure}
  \begin{center}
    \includegraphics[width=\columnwidth]{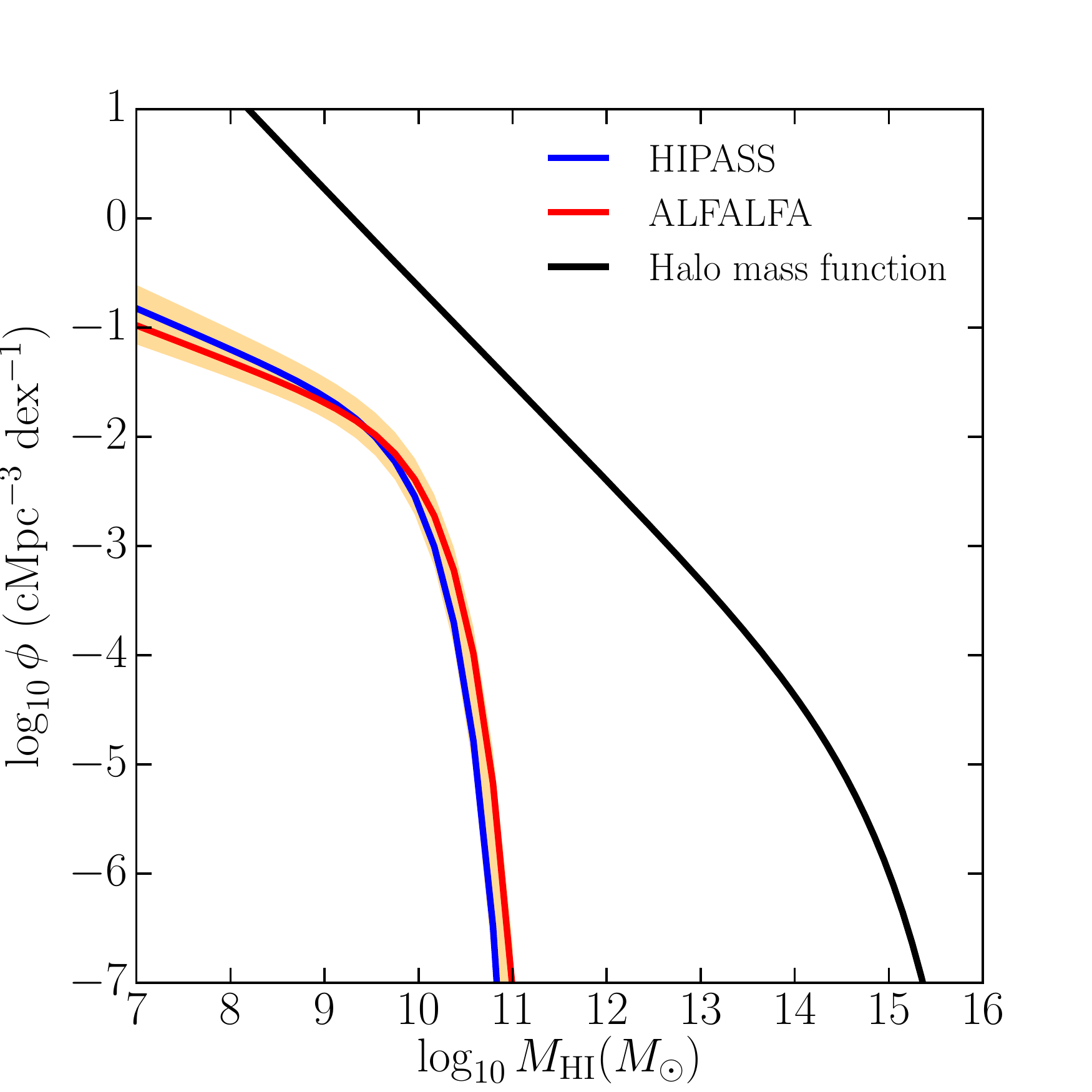}
  \end{center}
  \caption{The blue and red curves show the HI mass functions derived
    from the HIPASS \citep{zwaan05} and ALFALFA data \citep{martin10},
    respectively.  The shaded region shows the combined uncertainty.
    The black curve shows the halo mass function.}
  \label{fig:mhiandm}
\end{figure} 

We derive the HIHM relation at $z\sim 0$ by abundance matching dark
matter haloes with HI-selected galaxies.  We use the HI mass function
from the HIPASS \citep{meyer2004} and ALFALFA \citep{martin10} datasets,
the latter derived using the $1/V_{\rm max}$ as well as the 2DSWML
(2-Dimensional StepWise Maximum Likelihood) methods:

\begin{itemize}
\item HIPASS: This complete catalogue of HI sources contains 4,315
  galaxies \citep{meyer2004}. The HI mass function $\phi(M_{\rm HI})$ is fitted by a  Schechter function    using the the 2-Dimensional StepWise Maximum  Likelihood (2DSWML) method, with a total of 4010 galaxies.  The
  effective volume $V_{\rm eff}$ is calculated for each galaxy
  individually and the values of $1/V_{\rm eff}$ are summed in bins of
  HI mass to obtain the 2DSWML mass function.  The resultant best-fit
  parameters are $\alpha = -1.37 \pm 0.03 \pm 0.05$,
  $\log(M_{*}/M_{\odot}) = 9.80 \pm 0.03 \pm 0.03 h_{75}^{-2}$ and
  $\phi^* = (6.0 \pm 0.8 \pm 0.6) \times 10^{-3} h_{75}^3$ Mpc$^{-3}$
  (the two error values show statistical and systematic errors,
  respectively; \citealt{zwaan05}).  The distribution of HI masses is
  calculated using 30 equal-sized mass bins spanning $6.4 <
  \log_{10}M_{\rm HI} < 10.8$ (in $M_{\odot}$).

\item ALFALFA: This catalogue contains 10,119 sources to form the
  largest available sample of HI-selected galaxies \citep{martin10}.
  The ALFALFA survey measures the HI mass function by using both the
  2DSWML as well as the $1/V_{\rm max}$ methods.  The HI mass function
  is fitted with the Schechter form, with the best-fitting parameters
  $\phi^* = (4.8 \pm 0.3) \times 10^{-3} h_{70}^3$ Mpc$^{-3}$, $\log (M_*/M_{\odot}) + 2
  \log(h_{70}) = 9.95 \pm 0.04$, and $\alpha = -1.33 \pm 0.03$ with
  the $1/V_{\rm max}$ method, and $\phi^* = (4.8 \pm 0.3) \times
  10^{-3} h_{70}^3$ Mpc$^{-3}$, $\log (M_*/M_{\odot}) + 2 \log(h_{70}) = 9.96 \pm 0.2$,
  and $\alpha = -1.33 \pm 0.02$ with the 2DSWML method.  The two
  determinations of the HI mass function are in good agreement.\footnote{In the figures, we only indicate the ALFALFA 2DSWML mass function fit for clarity.}
\end{itemize}

To match HI-selected galaxies to dark matter haloes, we use the
Sheth-Tormen \citep{sheth2002} form of the dark matter halo mass
function.  Figure~\ref{fig:mhiandm} shows the comparison of the three
HI mass functions mentioned above with the halo mass function, which
is shown by the solid black curve.  This corresponds to the assumption
that each dark matter halo hosts one HI galaxy with its HI mass
proportional to the host dark matter halo mass.  The shaded region in
Figure~\ref{fig:mhiandm} shows the combined uncertainty in the
observed HI mass functions.  
Matching the abundance of the halo mass function and the fitted HI
mass function then leads to the relation between the HI mass and the
halo mass \citep[e.g.,][]{vale2004}:
\begin{equation}
  \int_{M (M_{\rm HI})}^{\infty} \frac{dn}{ d \log_{10} M'} \ d \log_{10} M' = \int_{M_{\rm HI}}^{\infty} \phi(M_{\rm HI}') \ d \log_{10} M_{\rm HI}'
  \label{eqn:abmatch}
\end{equation}
where $dn / d \log_{10} M$ is the number density of dark matter haloes with logarithmic
masses between $\log_{10} M$ and $\log_{10} (M$ + $dM)$, and $\phi(M_{\rm HI})$ is the
corresponding number density of HI galaxies in logarithmic mass bins.  Solving
Equation~(\ref{eqn:abmatch}) gives a relation between the HI-mass
$M_{\rm HI}$ and the halo mass $M$.  Note that this approach assumes
that there is a monotonic relationship between $M_{\rm HI}$ and $M$.

\begin{figure}
  \begin{center}
    \includegraphics[width=\columnwidth]{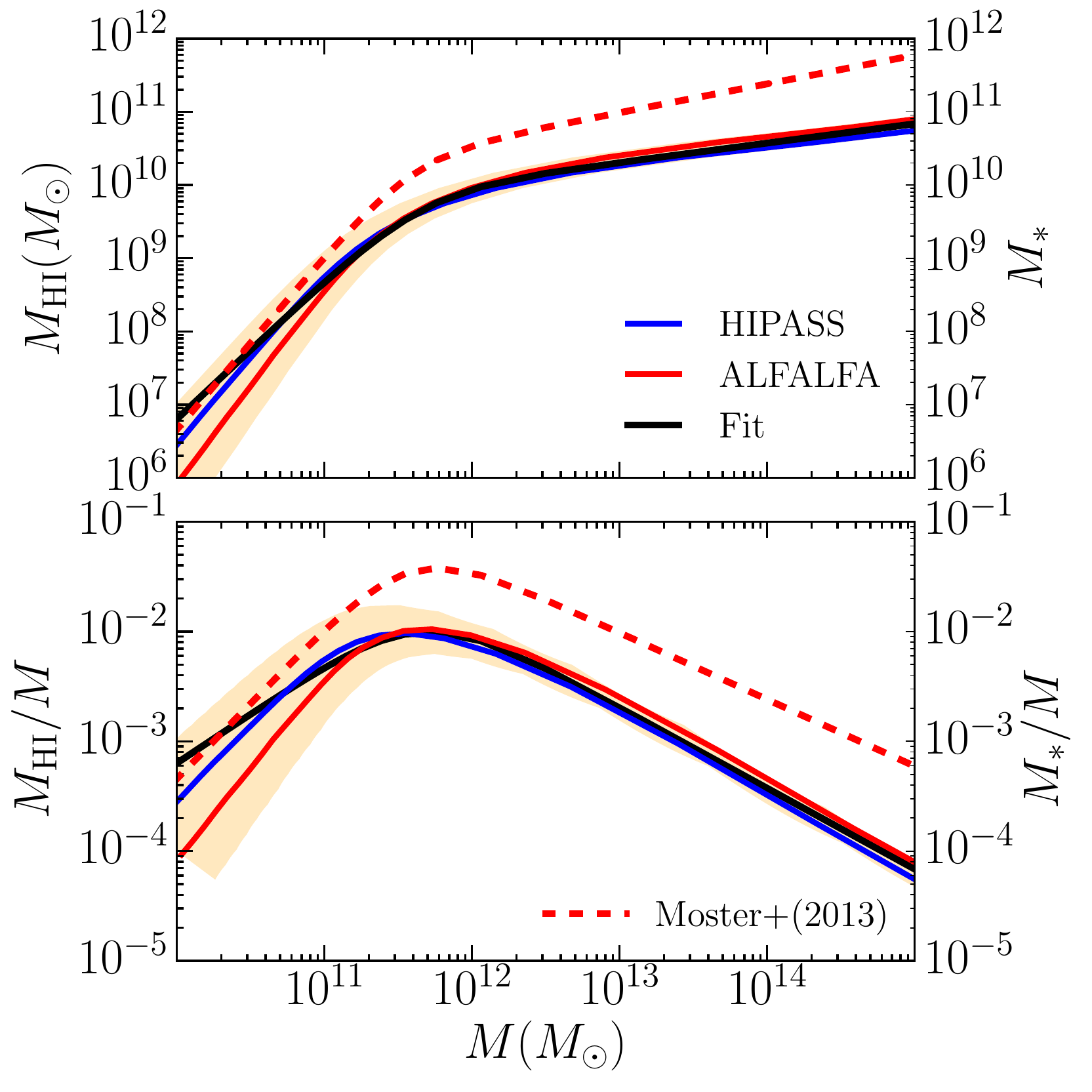}
  \end{center}
  \caption{\textit{Top panel}: The HIHM relation at $z=0$ derived from HIPASS
    (blue curve) and ALFALFA (red curve) HI mass functions.  The black
    curve shows a combined fit to the mass functions using the parametric form of
    Equation~(\ref{moster12}).  The shaded region shows the error in
    the fit.  \textit{Lower panel}: The HI mass fraction, $M_{\rm
      HI}/M$ as a function of halo mass $M$ at $z=0$.  Also shown for
    comparison in both panels is the SHM relation \citep{moster2013}.}
\label{fig:coldgasfrac}
\end{figure} 

Solving Equation~(\ref{eqn:abmatch}) in the mass range
$10^6$~M$_{\odot} < M_{\rm HI} < 10^{11}$~M$_{\odot}$, we show the
resultant HIHM relation in the top panel of
Figure~\ref{fig:coldgasfrac}.  The red curve shows the HIHM relation
obtained from the ALFALFA data, while the blue curve shown the same
for the HIPASS data.   We find that the HI mass monotonically increases
as a function of the halo mass and changes slope at a characteristic
value of the halo mass.  This behaviour is qualitatively similar to
the SHM relation \citep{moster2013}, which is shown by the dashed red
curve in the top panel of Figure~\ref{fig:coldgasfrac}.  For small
mass haloes, the HI mass is nearly equal to the stellar mass.  But the
HI mass decreases more rapidly than the stellar mass as a function of
halo mass, and for high mass haloes the HI mass is down to almost a tenth of
the stellar mass.  The characteristic mass for the HIHM relation is also slightly smaller ($10^{11.7} M_{\odot}$) than that for the SHM relation ($\sim 10^{12} M_{\odot}$).
The HIHM relation is shown as the ratio of the HI and halo masses in
the lower panel of Figure~\ref{fig:coldgasfrac}.  The peak HI mass
fraction is about 1\%, and  this reduces down to 0.01\% at both high and low
masses.  The peak HI mass fraction is in good agreement with the
abundance matching estimates of \citet{puebla2011, evoli2011,
  baldry2008} and the direct estimate of \citet{papastergis2012} for
the baryonic mass fraction. It had been found that the clustering of
the HI selected galaxies in ALFALFA \citep{papastergis2013} was also
well-matched by abundance matching at $z \sim 0$, and the cold gas
fraction showed a maximum at halo masses close to $10^{11.1 - 11.3}
M_{\odot}$, which was lower than the corresponding peak for the stellar mass
fraction ($10^{11.8} M_{\odot}$).

We parameterise the HIHM relation by a function of the form introduced
for the SHM relation by \citet{moster2013},
\begin{equation}
M_{\rm HI} = 2 N_{10} M \left[\left(\frac{M}{M_{10}}\right)^{-b_{10}} + \left(\frac{M}{M_{10}}\right)^{y_{10}}\right]^{-1}.
\label{moster12}
\end{equation}
We fit the HIHM relation by the function of this form using non-linear
least squares.  The best-fitting values of the free parameters are
$M_{10}=(4.58 \pm 0.19)\times 10^{11}$~M$_\odot$, $N_{10}=(9.89\pm
4.89)\times 10^{-3}$, $b_{10}=0.90 \pm 0.39$ and $y_{10}=0.74 \pm
0.03$.
The errors here are estimated by propagating the uncertainties in
Figure~\ref{fig:mhiandm}.  The best-fit HIHM relations are shown in
Figure~\ref{fig:coldgasfrac} (black curves), with the corresponding
error indicated by the shaded region.

\begin{figure}
  \begin{center}
    \includegraphics[width=\columnwidth]{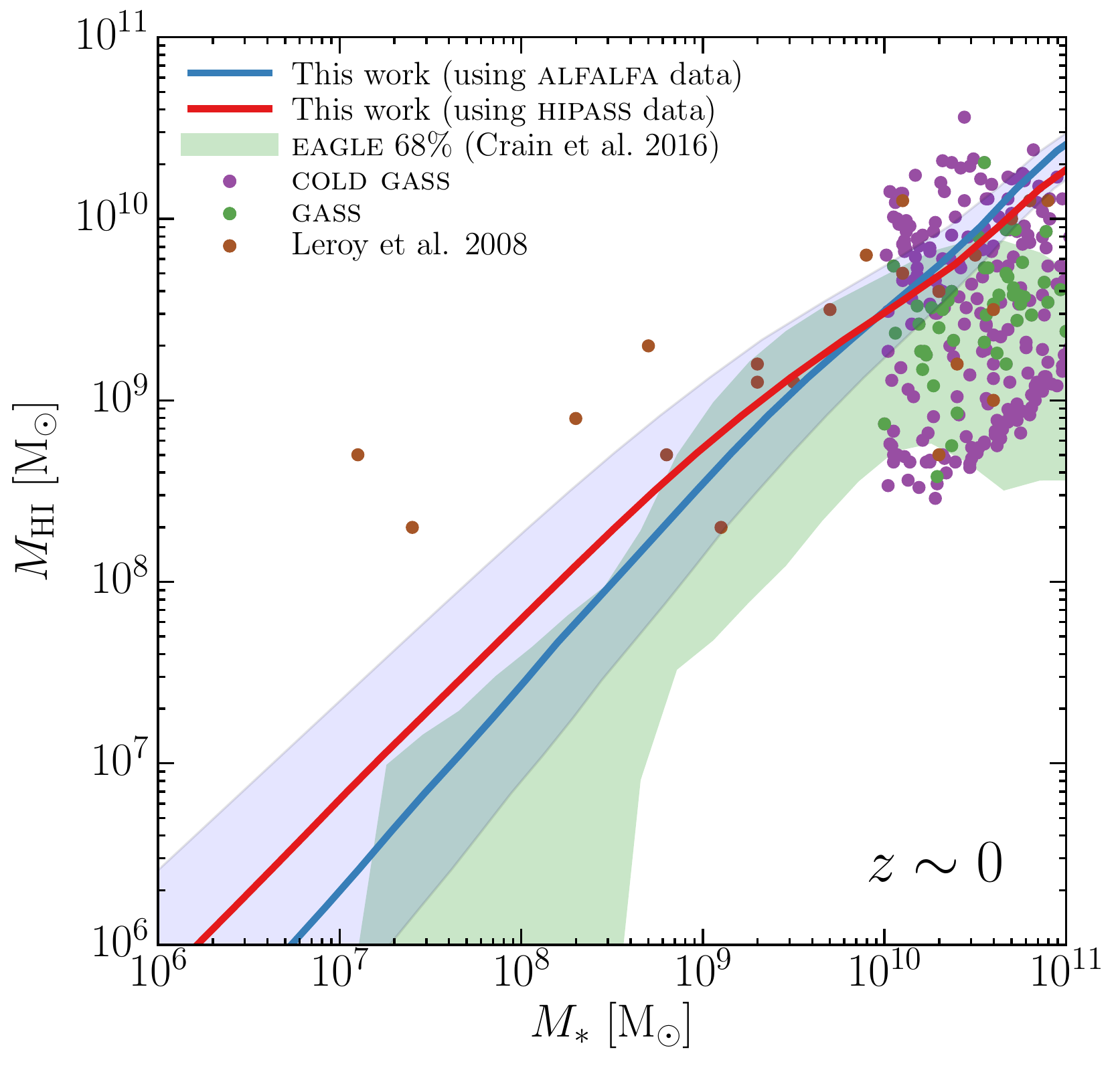} 
  \end{center}
  \caption{The HI-mass stellar-mass relation obtained by abundance
    matching combined with the SHM relation determined by
    \citet{moster2013}, are shown by the solid curves. {The 68\%
      scatter in the relation is indicated by the blue band.} The
    green band shows the region around the median in which 68\% of the
    galaxies in the EAGLE reference simulation lie on this plane
    \citep{eagle2016}. Also shown are the data from individual objects
    detected in the GASS and COLD GASS surveys, and the nearby
    galaxies in HERACLES and THINGS \citep{leroy2008}.}
  \label{fig:mhimstar}
\end{figure} 

\subsection{The HI-mass stellar-mass relation}

We can combine our derived HIHM relation with known SHM relations to
understand the relationship between the HI mass and stellar mass in
dark matter haloes.  \citet{moster2013} use a multi-epoch abundance
matching method with observed stellar mass functions (SMFs) to
describe the evolution of the SHM relation across redshifts.  At each
redshift, they parameterise the SHM relation using the functional form
in Equation~(\ref{moster12}).  At low redshifts, the SMFs of
\citet{li2009} based on the Sloan Digital Sky Survey (SDSS) DR7
\citep{york2000, abazajian2009} are used, along with the observations
of \citet{baldry2008}. At higher redshifts, the SMFs by
\citet{gonzalez2008} are used for massive galaxies, and those by
\citet{santini2012} for the low mass galaxies. From the results of
abundance matching, the mean SHM relation is obtained, which is then
used to populate haloes in the Millennium
\citep[MS-I;][]{millenium2005} and the Millennium - II
\citep[MS-II;][]{boylan2009} simulations with galaxies. From this, the
model stellar mass functions are derived and directly compared to
observations to constrain the free parameters in the SHM relation. The
resulting mean stellar mass fraction at $z \sim 0$ is shown by the
dashed line in Figure~\ref{fig:coldgasfrac}.

\begin{figure}
  \begin{center}
    \includegraphics[width=\columnwidth]{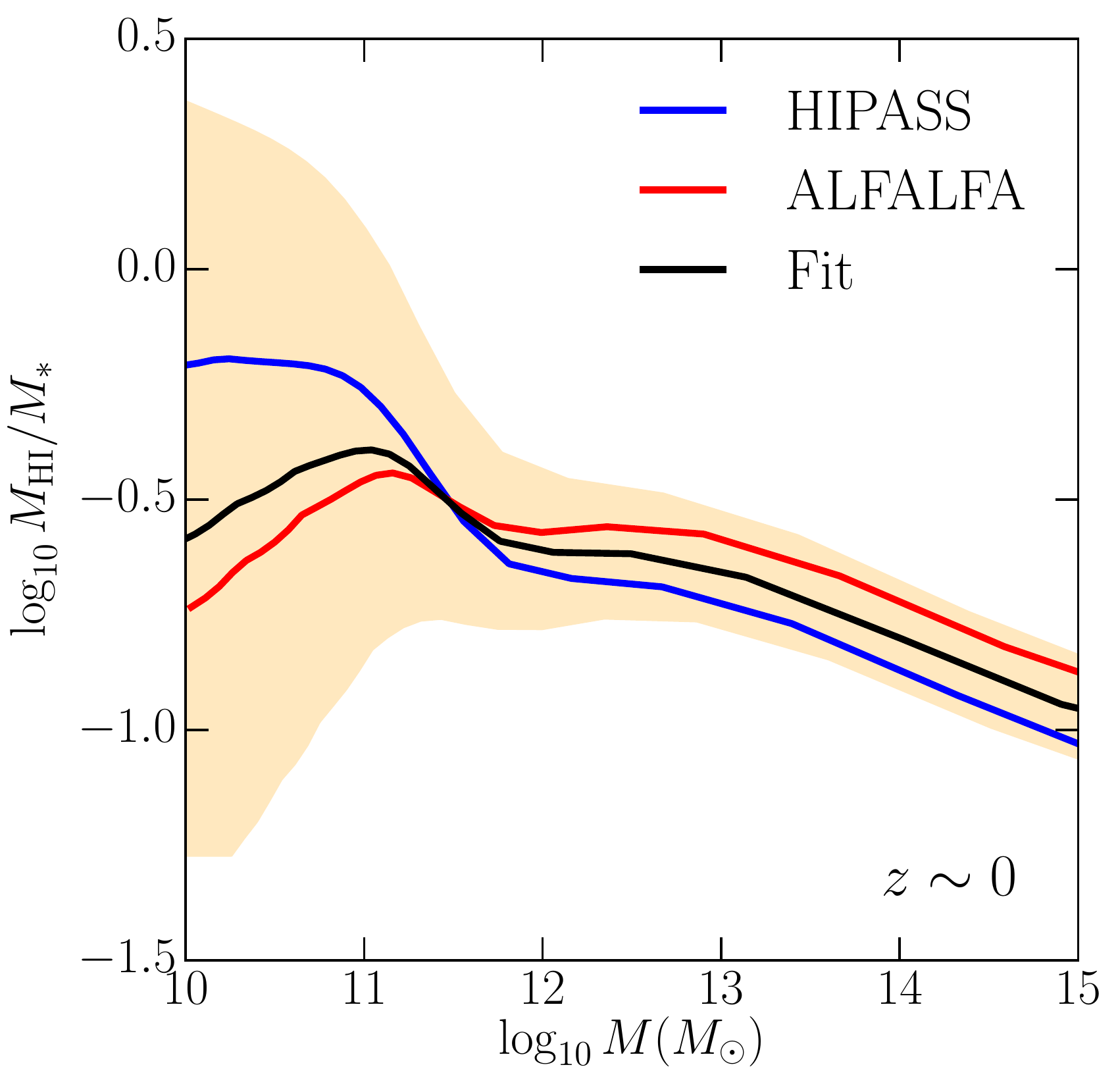} 
  \end{center}
  \caption{The HI-mass to stellar-mass ratio as a function of the halo
    mass at $z\sim 0$.  The blue and red curves combine our results
    for HIPASS and ALFALFA data, respectively, with the SHM relation
    from \citet{moster2013}. The parametrized fit is indicated by the black curve. The shaded region shows the uncertainty in the HI-mass to
    stellar-mass ratio obtained by propagating errors from
    Figure~\ref{fig:coldgasfrac}.}
  \label{fig:mhimstarm}
\end{figure} 

We use the \citet{moster2013} results for the SHM relation, coupled to
our abundance matching results for HIHM to arrive at a HI-mass
stellar-mass relation.  This is shown by the solid red and blue curves
in Figure~\ref{fig:mhimstar} for HIPASS and ALFALFA
respectively. {The 68\% scatter in the relation is indicated by
  the blue band.} For comparison, we also show the measurements from
750 galaxies in the redshift range $0.025 < z < 0.05$ and $M_{*} >
10^{10}$~M$_{\odot}$ from the GALEX Arecibo SDSS survey
\citep[GASS;][]{catinella2010, catinella2013}, and 366 galaxies from
the COLD GASS survey \citep{saintonge2011, saintonge2011a,
  catinella2012}.  We also show results from \citet{leroy2008}, which
is a compilation of individual galaxies detected in the HERA CO Line
Extragalactic Survey \citep[HERACLES;][]{leroy2009} that are part of
The HI Nearby Galaxy Survey \citep[THINGS;][]{walter2008}, which
covers \HI{} masses in the range $(0.01$--$14) \times
10^9$~M$_{\odot}$.  These measurements are consistent with our result,
although the observational data exhibit a somewhat large
scatter. {We note that the HI-stellar mass relation
    from the ALFALFA data and the THINGS data show some discrepancy at
    low stellar masses (also seen in \citet{popping2015}, which
    matches the data in \citet{leroy2008}, but has difficulty matching
    the ALFALFA data mass function at low HI
    masses).  However, the main
    aim of the present work is to provide an understanding of the
    HI-mass halo-mass relation, and as such, we do not conjecture on
    the observed discrepancy of the \citet{leroy2008} results with the
    ALFALFA data.} We also compare our HI-mass stellar-mass relation
with that found in the EAGLE hydrodynamical simulations
\citep{schaye2015, crain2015}.  The EAGLE simulations model the
formation and evolution of galaxies in the presence of various
feedback processes.  They also model the HI content of galaxies by
using calibrated fitting functions from radiative transfer simulations
to estimate self-shielding, and also employing empirical relations to
correct for molecular gas formation \citep{eagle2016}. The green band
in Figure~\ref{fig:mhimstar} shows the region around the median on the
HI-mass stellar-mass diagram occupied by 68\% of galaxies in the
reference EAGLE simulation (labelled ``L100N1504'' in
\citealt{schaye2015}).  Our results are in good agreement with the
EAGLE predictions, except possibly at the highest stellar masses
($M_*>10^{10}$~M$_\odot$) where the HI mass in EAGLE galaxies starts
to decrease.  This is likely a reflection of the AGN feedback in
EAGLE, that heats and expunges cold gas from high mass galaxies by
their massive central black holes \citep{eagle2016}.

Figure~\ref{fig:mhimstarm} shows the HI-mass to stellar-mass ratio as
a function of the halo mass.  The blue and red curves show the results
for HIPASS and ALFALFA respectively, and the black curve shows the
parametrized fit.  In each case, we obtain the HI-mass to stellar-mass
ratio by combining our HIHM relation with the SHM relation of
\citet{moster2013}.  The HI-mass to stellar-mass ratio is about 25\%
in a rather broad range of halo masses from $10^{11}$ to
$10^{13}$~M$_\odot$.  {The ratio decreases to about 10\% at halo masses
above this range, and is more uncertain below this range, due to the uncertainty in the data and the fitting (Fig. \ref{fig:coldgasfrac} lower panel) at lower masses.}  The shaded regions show the
uncertainty in the HI-mass to stellar-mass ratio, obtained by
propagating the errors from Figure~\ref{fig:coldgasfrac}. 

\section{HIHM relation at high redshift}
\label{sec:observations}

\begin{table}
\centering
\begin{tabular}{cll}
\hline
$z$   & Observable               & Source                  \\
\hline
$\sim$ 1    & $\Omega_{\rm HI}b_{\rm HI}$ & \citet{switzer13}       \\
     & $f_{\rm HI}$               & \citet{rao06}           \\
     & $dN/dX$                   & \citet{rao06}           \\
2.3  & $\Omega_{\rm DLA}$         & \citet{zafar2013}       \\
     & $f_{\rm HI}$               & \citet{noterdaeme12}    \\
     & $b_{\rm DLA}$              & \citet{fontribera2012}  \\
     & $dN/dX$                   & \citet{zafar2013}       \\
 > 3 & $dN/dX$                   & \citet{zafar2013}       \\
 \hline
\end{tabular}
\caption{High-redshift data used in this paper.  The measurement of
  $\Omega_{\rm HI}b_{\rm HI}$ comes from HI intensity mapping at
  $z\sim 0.8$ by \citet{switzer13}. \citet{rao06} use measurements of
  absorption systems at median redshifts $z \sim 0.609$ and $z \sim
  1.219$ to derive the DLA parameters. All other data come from
  Lyman-$\alpha$ absorption measurements using high-redshift quasar
  spectra.}
\label{table:data}
\end{table}

Due to the intrinsic faintness of the 21~cm line, the direct detection
of HI from resolved galaxies is difficult at redshifts above $z \sim
0.1$.  At higher redshifts ($z<5$), therefore, constraints on the
distribution and evolution of HI in galaxies mainly come from high
column density Lyman-$\alpha$ absorption systems (Damped
Lyman-$\alpha$ Absorbers; DLAs) with column densities
$N_\mathrm{HI}>10^{20.3}$~cm$^{-2}$ in the spectra of bright
background sources such as quasars.  The relevant observables at these
redshifts are the incidence rate $dN/dX$ of DLAs, the column density
distribution $f_\mathrm{HI}(N_\mathrm{HI},z)$ of DLAs at high column
densities, the three-dimensional clustering of DLAs as quantified by
their clustering bias relative to the underlying dark matter, and the
total amount of neutral hydrogen in DLAs \citep{wolfe1986,
  lanzetta1991, gardner1997, prochaska09, rao06, noterdaeme12,
  zafar2013}.  A detailed summary of the low- and high-redshift HI
observables is provided in \citet{hptrcar2015}. We now extend the HIHM
relation obtained at $z=0$ to higher redshifts by using these
observables. Throughout the analysis, we use the cosmological
parameters $h = 0.71$, $\Omega_m = 0.281$, $\Omega_{\Lambda} = 0.719$,
$\sigma_8 = 0.8$, $n_s = 0.964$. 

\subsection{Modelling the HI observables}

To model the distribution of HI density within individual dark matter
haloes, we use the redshift- and mass-dependent modified
Navarro-Frenk-White (NFW; \citealt{1996ApJ...462..563N}) profile
introduced by \citet{barnes2014}:
\begin{equation}
  \rho_{\rm HI}(r) = \frac{\rho_0 r_s^3}{(r + 0.75 r_s) (r+r_s)^2},
  \label{rhodef}
\end{equation} 
where $r_s$ is the scale radius defined as $r_s=R_v(M)/c(M,z)$, with
$R_v(M)$ being the virial radius of the halo.  The halo concentration
parameter, $c(M,z)$ is approximated by:
\begin{equation}
  c(M,z) = c_{\rm HI} \left(\frac{M}{10^{11} M_{\odot}} \right)^{-0.109} \left(\frac{4}{1+z} \right).
\end{equation} 
The profile in Equation~(\ref{rhodef}) is motivated by {the analytical modelling of} cooling in multiphase halo gas by \cite{maller2004}.  In the above
equation, $c_{\rm HI}$ is a free parameter, the concentration
parameter for the HI, analogous to the dark matter halo concentration
$c_0 = 3.4$ \citep{maccio2007}. The value of this parameter can be
constrained by fitting to the observations.  The 
$\rho_0$ in Equation~(\ref{rhodef}) is determined by normalization to
the total HI mass:
\begin{equation}
 \int_0^{R_v(M)} 4 \pi r^2 \rho_{\rm HI}(r) dr = M_{\rm HI} (M)
\end{equation} 
Thus, both the HI-halo mass relation as well as the radial
distribution of HI are required for constraining the HI profile.

\begin{figure}
  \begin{center}
    \includegraphics[width = \columnwidth, scale=0.45]{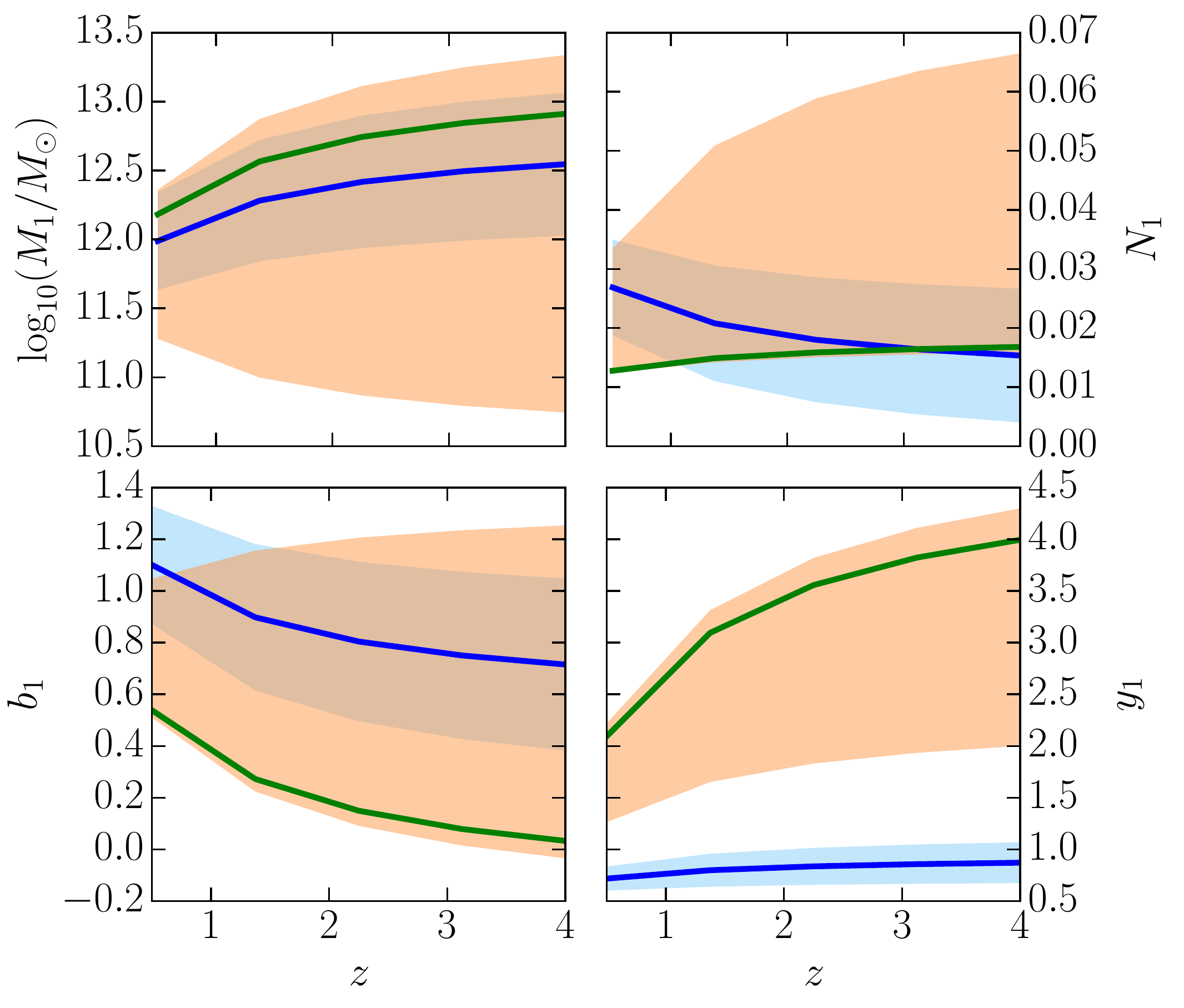} 
  \end{center}
  \caption{The evolution of the parameters of the HIHM relation
    (Equation~\ref{mosterredshiftevol}).  The green curves show our
    best-fit parameter inferences with 68\% confidence intervals shown
    by the orange shaded region. For comparison, the evolution of the
    corresponding quantities for the SHM relation of
    \citet{moster2013} is shown in blue.}
  \label{fig:evolution}
\end{figure}

 \begin{figure*}
  \begin{center}
    \includegraphics[width=\textwidth]{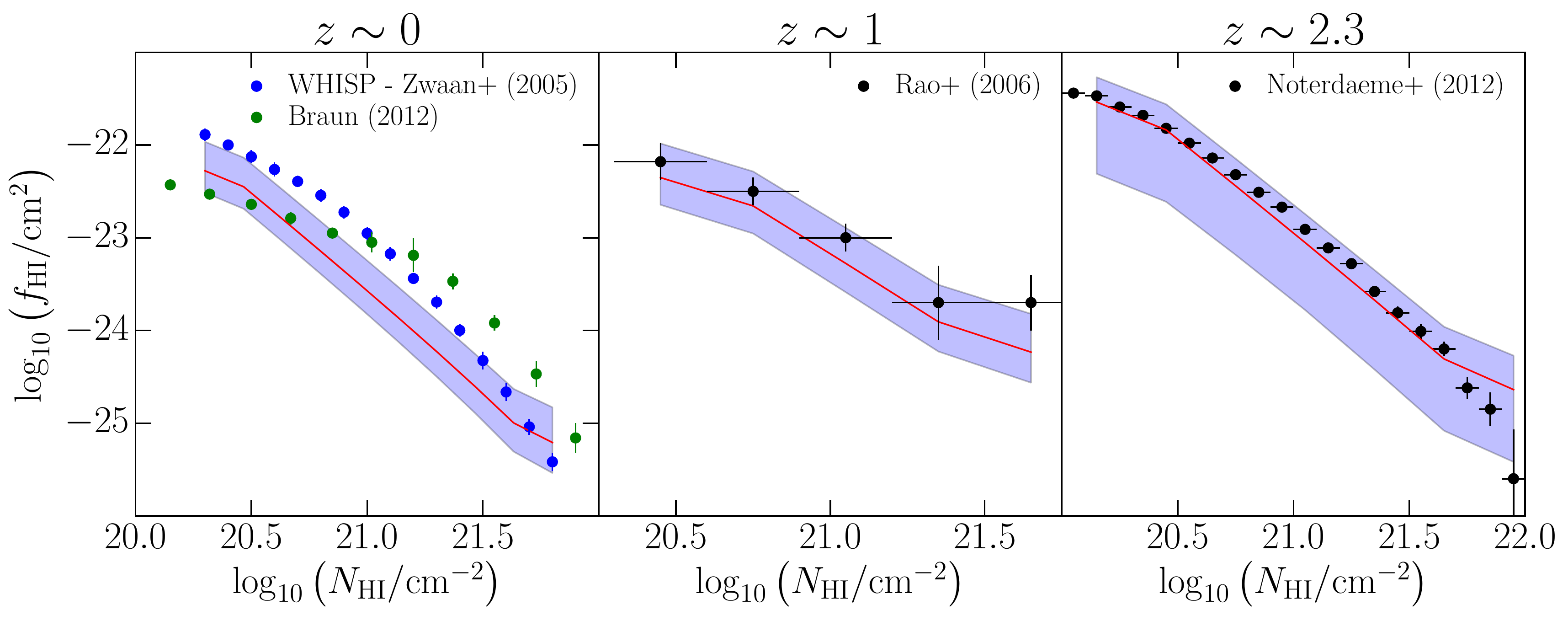} 
  \end{center}
  \caption{The best-fit column density distribution (red curves) in
    our model at redshifts 0, 1 and 2.3, compared to the
    observations. The blue shaded regions show the 68\% confidence
    limits.  The model fits the high redshift column density
    distributions quite well but has difficulty in fitting the column
    density distribution at $z=0$, especially at low column
    densities.}
  \label{fig:fhibias0}
\end{figure*}  

The DLA based quantities at different redshifts can now be computed by
defining the column density of a halo at impact parameter $s$ as
\citep{barnes2014, hptrcar2016}:
\begin{equation}
 N_{\rm HI}(s) = \frac{2}{m_H} \int_0^{\sqrt{R_v(M)^2 - s^2}}dl \ \rho_{\rm HI}\left(\sqrt{s^2 + l^2}\right)
 \label{coldenss}
\end{equation} 
where $m_H$ is the hydrogen atom mass and $R_v(M)$ is the virial radius
associated with a dark matter halo of mass $M$.  We define the DLA
cross-section of the halo as $\sigma_{\rm DLA} = \pi s_*^2$, where
$s_*$ is defined such that $N_{\rm HI}(s_*) = 10^{20.3}$ cm$^{-2}$.  The
clustering bias of DLAs, $b_{\rm DLA}$, can then be written as 
\begin{equation}
 b_{\rm DLA} (z) =  \frac{\int_{0}^{\infty} dM n (M,z) b(M,z) \sigma_{\rm DLA} (M,z)}{\int_{0}^{\infty} dM n (M,z) \sigma_{\rm DLA} (M,z)},
\end{equation}
where $n(M,z)$ is the comoving halo mass function and $b(M,z)$ is the
clustering bias factor of haloes \citet{scoccimarro2001}.  The DLA
incidence $dN/dX$ can be calculated as
\begin{equation}
 \frac{dN}{dX} = \frac{c}{H_0} \int_0^{\infty} n(M,z) \sigma_{\rm DLA}(M,z) \ dM,
 \label{dndxdef}
\end{equation} 
and the column density distribution $f_{\rm HI}(N_{\rm HI}, z)$ is given by
\begin{multline}
  f(N_{\rm HI}, z) \equiv \frac{d^2 n}{dX d N_{\rm HI}} \\
  = \frac{c}{H_0} \int_0^{\infty} n(M,z) \left|\frac{d \sigma}{d N_{\rm HI}} (M,z) \right| \ dM
  \label{coldensdef}
\end{multline}
where
\begin{equation}
  \frac{d\sigma}{dN_{\rm HI}}=2\pi s\frac{ds}{dN_{\rm HI}},
\end{equation}
with $N_{\rm HI}(s)$ defined by Equation~(\ref{coldenss}).  The
density parameter for DLAs, $\Omega_{\rm DLA}$ is obtained by
integrating the column density distribution 
\begin{equation}
 \Omega_{\rm DLA}(N_{\rm HI}, z)  = \frac{m_H H_0}{c \rho_{c,0}} \int_{10^{20.3}}^{\infty} f_{\rm HI}(N_{\rm HI}, z) \ N_{\rm HI} \ d N_{\rm HI},
\end{equation}
where $\rho_{c,0}$ is the present-day critical density.

At high redshifts, we also use the measurement of $\Omega_{\rm HI}b_{\rm
  HI}$ from HI intensity mapping at $z\sim 0.8$ by \citet{switzer13}.
To calculate this quantity in our model, the HI density parameter is
given by
\begin{equation}
 \Omega_{\rm HI} (z) = \frac{1}{\rho_{c,0}} \int_0^{\infty} n(M, z) M_{\rm HI} (M,z) dM \ .
 \label{omegaHI}
\end{equation} 
The bias of HI is given by
\begin{equation}
b_{\rm HI} (z) = \frac{\int_{0}^{\infty} dM n(M,z) b (M,z) M_{\rm HI} (M,z)}{\int_{0}^{\infty} dM n(M,z) M_{\rm HI} (M,z)}
\label{biasHI}
\end{equation}
where $b(M,z)$ is the dark matter halo bias. We fit the HI density
profiles of haloes at $z=0$ by using the  column
density distribution at $z=0$ for $N_\mathrm{HI}>10^{20.3}$~cm$^{-2}$,
derived from the WHISP data by \citet{zwaan2005a}.

\subsection{Extending the HIHM relation to high redshifts}
\label{sec:extending}
We can now extend the HIHM relation developed in
Section~\ref{sec:abmatch} to higher redshifts.  We do this by
parameterising the HIHM relation evolution in a manner similar to the
parameterisation of the SHM relation evolution by \citet{moster2013}.
We write the HIHM relation at higher redshifts as
\begin{equation}
M_{\rm HI} = 2 N_{1} M \left[\left(\frac{M}{M_{1}}\right)^{-b_{1}} + \left(\frac{M}{M_{1}}\right)^{y_{1}}\right]^{-1},
\label{mosterredshiftevol}
\end{equation}
which has the same form as Equation~(\ref{moster12}).  The parameters
in Equation~(\ref{mosterredshiftevol}) are written as:
\begin{align}
& \log_{10} M_{1} = \log_{10}  M_{10} + \frac{z}{z + 1}  M_{11}, \nonumber \\
& N_{1}  = N_{10} + \frac{z}{z + 1} N_{11}, \nonumber \\ 
& b_{1} = b_{10} + \frac{z}{z + 1} b_{11},\ \mathrm{and}\nonumber \\
& y_{1}  = y_{10} + \frac{z}{z + 1} y_{11}. 
\label{eq:evol}
\end{align}

The parameters $M_{10}$, $N_{10}$, $b_{10}$ and $y_{10}$ are defined
in Equation~(\ref{moster12}) for $z=0$.  The four additional
parameters, $M_{11}$, $N_{11}$, $b_{11}$ and $y_{11}$, introduced by
Equations~(\ref{eq:evol}) govern the evolution of the HIHM at high
redshift.  These four parameters together with the HI density profile
parameter $c_{\rm HI}$ are to be constrained from the high redshift
observations.  This is done by using the data available from $z=0$ to
$5$ as summarised in Table~\ref{table:data}.  We use the measurements  of the incidence rate $dN/dX$ of DLAs, the column
density distribution $f_\mathrm{HI}(N_\mathrm{HI},z)$ of DLAs at high
column densities, the three-dimensional clustering of DLAs as
quantified by their clustering bias relative to the dark matter,
and the total amount of neutral hydrogen in DLAs \citep{wolfe1986,
  lanzetta1991, gardner1997, prochaska09, rao06, noterdaeme12,
  zafar2013}, as well as the measurements of the HI column density
distribution and clustering from radio data at $z<1$ \citep{zwaan2005a,
  switzer13}.

\begin{figure}
  \begin{center}
    \hskip-0.2in \includegraphics[width = \columnwidth, scale=0.6]{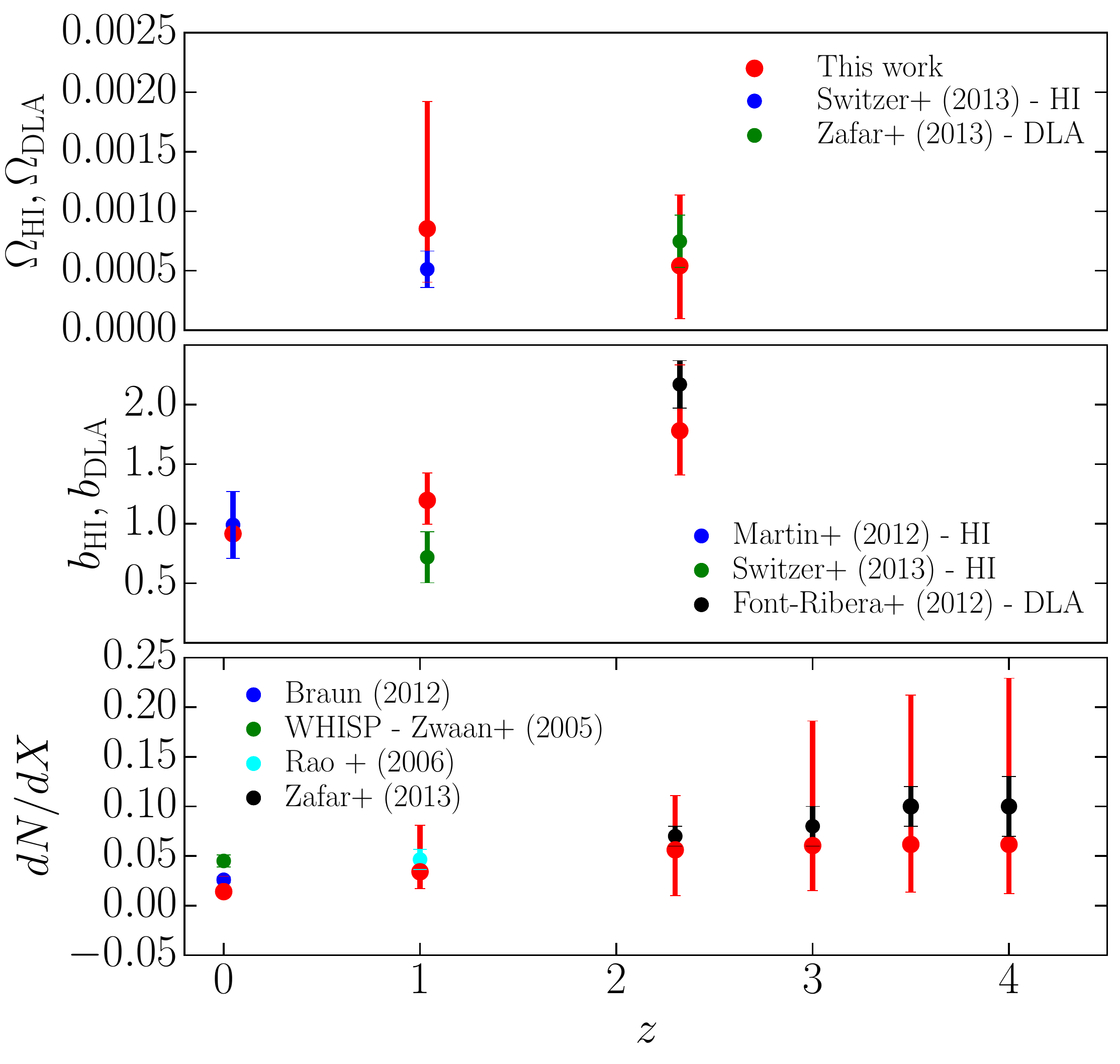} 
  \end{center}
\caption{Our model predictions for the density parameter, clustering
  bias, and DLA incidence rate (red, with 68\% confidence intervals
  indicated by the error bars) compared to the observations.  Note
  that at redshift $z \sim 1$, \citet{switzer13} constrain the product
  $\Omega_{\rm HI} b_{\rm HI}$.  Shown here is the observed
  $\Omega_{\rm HI} b_{\rm HI}$ divided by the model value of $b_{\rm
    HI}$ (top panel) and $\Omega_{\rm HI}$ (second panel).  The model
  successfully matches these observations, including the bias at high
  redshifts.}
\label{fig:panels123}
\end{figure} 

The best-fitting values for the five parameters $M_{11}, N_{11},
b_{11}$, $y_{11}$ and $c_{\rm HI}$, and their errors are now estimated
by a Bayesian Markov Chain Monte Carlo (MCMC) analysis using the
\textsc{CosmoHammer} package \citep{akeret2013}. The likelihood,
\begin{equation}
\mathcal{L} = \exp\left(-\frac{\chi^2}{2}\right) 
\end{equation}
is maximized with respect to the five free parameters, with:
\begin{equation}
\chi^2 = \sum_i\frac{(f_{\rm i} - f_{\rm obs,i})^2}{\sigma^2_{\rm obs,i}}
\end{equation}
where the $f_{\rm i}$ are the model predictions, $f_{\rm obs,i}$ are
the observational data and $\sigma^2_{\rm obs,i}$ are the squares of
the associated uncertainties (here assumed independent).

The best fitting parameters and their 68\% errors are
$M_{11}=1.56^{+0.53}_{-2.70}$,
$N_{11}=0.009^{+0.06}_{-0.001}$, $b_{11}=-1.08^{+1.52}_{-0.08}$,
$y_{11}=4.07^{+0.39}_{-2.49}$, and $c_{\rm
  HI}=133.66^{+81.39}_{-56.23}$.  The inferred evolution of the four
parameters of the HIHM relation in Equation~(\ref{mosterredshiftevol})
is shown in Figure~\ref{fig:evolution} together with the 68\%
errors.  For comparison, the evolution of the corresponding parameters
in the SHM relation parametrization of \citep{moster2013} are also
shown.  {The model allows for a wide range of parameters in the HIHM relation at high redshifts.}  The increase in the {best-fitting} characteristic mass follows the
increase in the characteristic halo mass of the SHM relation.  The
evolution of the high mass slope $y_1$ is much more rapid for the HIHM
relation than the SHM relation.  As we will see below, the high value
of the clustering bias factor for DLAs at high redshifts forces the
increase in the characteristic halo mass of the HIHM relation but the
more gradual increase observed in the DLA incidence rate prevents us
from putting too much HI in high mass halos, which constrains the high
mass slope to very steep values. 

Figure~\ref{fig:fhibias0} shows the column density distribution
derived from our model at $z\sim 0, 1,$ and $2.3$ together with the
associated 68\% statistical error. 

{At $z \sim 0$, only the concentration parameter of the profile is used to obtain the column density distribution, since the  HIHM relation has been directly fixed by the results of abundance matching. The concentration parameter is assumed to be equal to that obtained from the fitting of higher redshifts, which is done using the analysis outlined in Sec. \ref{sec:extending}.} The relation fits the available
data reasonably well, but leads to an underprediction of the observed
column density distribution at $z \sim 0$ at low column densities
($N_\mathrm{HI}<10^{21.4}$~cm$^{-2}$).  \footnote{{The two datasets for the column density distribution at $z \sim 0$ (which indicate a systematic offset) are shown only for comparison, and not directly fitted. The parameters involved in the HIHM are obtained from the abundance matching fits, and the concentration parameter is obtained from the results of the higher-redshift column density fitting. 
 The steep
slope of the HIHM relation  for $z=0$ leads to a lower column
density distribution than observed, suggesting that {the
  altered NFW profile may not fully describe} the HI density profiles
of halos at $z=0$, or that there {may be} a possible tension between
the HI mass function and the column density distribution at $z=0$. We
explore this issue in further detail in future work.}} Figure~\ref{fig:panels123}
compares other quantities in our model to their observed values.  The
incidence rate of DLAs is fit very well by the model throughout the
redshift range considered here.  {The measurements of the density
parameters of HI and DLAs, and the clustering bias of $z \sim 2.3$ DLAs are also fit well.}  The fit to the measured HI bias at $z=0$ is also good,
although it is somewhat poor at $z=1$.

\begin{figure*}
  \begin{center}
    \includegraphics[scale=0.6, width = \textwidth]{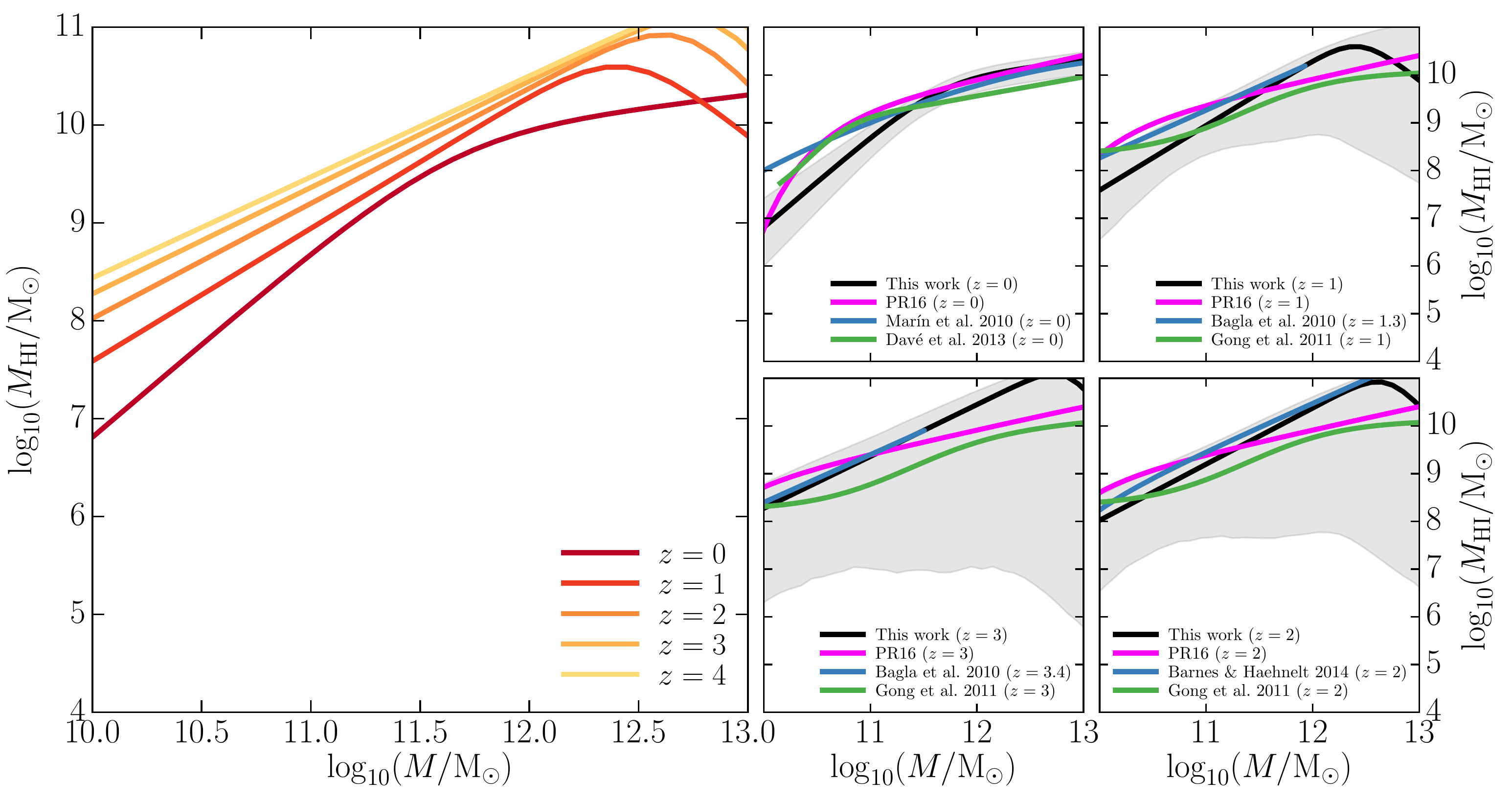} 
  \end{center}
  \caption{\textit{Left panel}: The HIHM relation inferred at redshifts $z=0$,
    $1$, $2$, $3$ and $4$ from the present work. \textit{Right panels}: The
    HIHM relation relation in the present work compared to the results
    of other approaches in the literature at redshifts $z=0$, $1$, $2$
    and $3$.}
  \label{fig:massevolall}
\end{figure*} 
 
\subsection{Comparison to other models of HI at high redshift}

Figure~\ref{fig:massevolall} shows the inferred best-fitting HIHM at $z=0$, $1$,
$2$, $3$ and $4$ in the present model, together with their associated uncertainties. In each case, the black curve shows the best-fit HIHM relation
  and the grey band shows the 68\% scatter around it. The figure also presents a
comparison of the HIHM obtained from hydrodynamical simulations and
other approaches in the literature at $z=0$, $1$, $2$ and $3$. {{These are briefly described below:
\begin{enumerate}

\item  At $z=0$, the model that comes closest to the present work is the
non-parametric HIHM relation of \citet{marin2010}, although their
low-mass slope is shallower.  

\item The hydrodynamical simulations of
\citet{dave2013} produce an HIHM relation that has very similar
high-mass and low-mass slopes as the present HIHM relation.  The high characteristic mass of
the {average best-fitting} HIHM relation in the present work is a natural consequence of
matching the abundance of haloes with HI-selected galaxies, under the
assumption that HI-mass of dark matter haloes scales monotonically
with their virial mass.

\item  \citet{bagla2010} used a set of analytical
prescriptions to populate \HI{} in dark matter haloes.  In their
simplest model, \HI{} was assigned to dark matter haloes with a
constant fraction $f$ by mass, within a mass range. The maximum and
minimum masses of haloes that host \HI{} were assumed to be
redshift-dependent. It was also assumed that haloes with virial
velocities of greater than 200 km/s and less than 30 km/s do not host
any HI.  

\item \citet{gong2011} provide nonlinear analytical forms of the
HIHM relation at $z=1$, $2$ and $3$, derived from the results of the
simulations of \citet{obreschkow2009a}. These predict a
  slightly different form for the HIHM relation. 
  
  \item  The model of
\citet{barnes2014} uses an HIHM relation that reproduces the observed
bias of DLA systems at $z \sim 2.3$, and constrains stellar feedback
in shallow potential wells. 

\item \citet{2016arXiv160701021P} used a {statistical data-driven} approach to
derive the best-fitting HIHM relation and radial distribution profile
$\rho_{\rm HI}(r)$ for $z=0$--$4$, from a joint analysis combining the
data from the radio observations at low redshifts and the Damped
Lyman-Alpha (DLA) system observables at high redshifts, along the
lines of the present work. This approach also produces results
  consistent with the present work, although the present best-fit HIHM
  relation at high redshifts may prefer a higher characteristic halo mass.
  
  \end{enumerate}

  It can be seen that all these
  models are consistent with each other and with the data at the 68\% confidence level. Tighter constraints on the HIHM relation at high redshifts may be achieved with the availability of better quality data with upcoming radio telescopes.

}}

\section{Conclusions}
\label{sec:conc}

In this paper, we have explored the evolution of the neutral hydrogen
content of galaxies in the last 12 Gyr (redshifts $z=0$--$4$).  At
redshift $z=0$, this work follows the approach of abundance matching,
which has been widely used for the stellar mass content of galaxies to
model galaxy luminosity functions \citep{vale2004,
  vale2006, conroy2006, shankar2006,
  guo2010, behroozi2010, moster2010, moster2013}.
A parameterised functional form for a monotonic relationship between
the \HI{} and halo mass is assumed to obtain the HI- Halo Mass (HIHM)
relation.  The best fit values of the parameters that fit the observed
\HI{} mass function from radio data are then obtained.  This approach
of modelling the HIHM relation at $z=0$ from the radio data at low
redshifts has been followed previously by \citet{papastergis2013}.
Our abundance matched HIHM agrees with that derived by these authors.

We further explore how well the abundance matching approach at $z = 0$
can be constrained by fitting to the high redshift data. We extend the
low redshift determination of the HIHM relation by postulating that
the evolution of the HIHM relation is similar to the
stellar-to-halo-mass (SHM) relation.  We parameterize this evolution
analogously to the evolution of the SHM relation by
\citet{moster2013}. {The physical motivation for the
  parametrization is that the HI-follows-stars functional form works
  well at low redshifts, which is in turn a consequence of the fact
  that the underlying mass/luminosity functions can both be described
  by the Schechter form.} Observational measurements of the \HI{} mass
function are not {yet} available at these redshifts.  Hence, we use
measurements of the \HI{} column density distribution function and the
\HI{} clustering from UV/optical observations of quasar absorption
spectra.  We assume that high column density systems (DLAs;
$N_\mathrm{HI}>10^{20.3}$~cm$^{-2}$) probe systems are high-redshift
analogs of \HI{} in galaxies detected in radio surveys at low
redshifts \citep{zwaan2005a}.

Our procedure allows a modeling of low and high redshift measurements
of the \HI{} content of galaxies to obtain the evolution of the HIHM
relation from $z=0$ to $2.3$ with the associated uncertainty.  This
technique is complementary to the forward modelling approach which
aims to characterize \HI{} using a halo model framework similar to
that of the underlying dark matter \citep{2016arXiv160701021P}.
{However, the present work represents a first attempt to characterize
  the HIHM relation empirically, directly from the data. Due to the
  sparse nature of the high-redshift data at present, there is
  considerable scatter in the high-redshift HIHM relation. {As
    a result,  other apparently dissimilar models from the
    literature are also consistent with the data and the allowed range of the present work. The scatter in the HIHM relation at higher redshifts can be reduced with tighter constraints on the HI mass functions from upcoming and future radio surveys.} 

Our results provide a useful benchmark to calibrate the \HI{} physics
in hydrodynamical simulations, especially at low redshifts where
correct treatment of star formation and feedback as well as cooling
and formation of molecular hydrogen are critical. They also provide an
estimate of the uncertainty in the HIHM relation coming from the
high-redshift data, and motivate further work towards possibly tighter
constraints on the HIHM relation.

\section*{Acknowledgements}

We thank Alireza Rahmati, Alexandre Refregier and Sergey Koposov for
useful discussions, Daniel Lenz for pointing out a minor typo and Robert Crain for kindly providing the data
from the EAGLE simulations. This work has made use of the VizieR
catalogue access tool, CDS, Strasbourg, France. The original
description of the VizieR service was published in the A\&AS 143, 23.
HP's research is supported by the Tomalla Foundation.  GK gratefully
acknowledges support from the ERC Advanced Grant 320596 `The
Emergence of Structure During the Epoch of Reionization'.

\bibliographystyle{mnras}
\bibliography{mybib}

\bsp
\label{lastpage}
\end{document}